# Preliminary results concerning the fracture-induced electromagnetic emissions recorded prior to the Mw 6.7 Earthquake on October 25, 2018 south of Zakynthos (Greece).

**S. M. Potirakis[1], Y. Contoyiannis[1], N. S. Melis[2], G. Koulouras[1], J. Kopanas[3], G. Antonopoulos[3], K. Eftaxias[3], C. Nomicos[1].**

1. Department of Electrical and Electronics Engineering, University of West Attica, Campus 2, 250 Thivon and P. Ralli, Aigaleo, Athens, GR-12244, Greece (S.M.P.: spoti@uniwa.gr; Y.C: yiaconto@uniwa.gr; G.K: gregkoul@uniwa.gr; C.N.: cnomicos@uniwa.gr)
2. Institute of Geodynamics, National Observatory of Athens, Lofos Nimfon, Thissio, Athens, GR-11810, Greece, nmelis@noa.gr
3. Department of Physics, Section of Solid State Physics, University of Athens, Panepistimiopolis, GR-15784, Zografos, Athens, Greece, (J.K.: jkopan@otenet.gr; G.A.: sv8rx@teiion.gr; K.E: ceftax@phys.uoa.gr)

**Abstract**

In this letter we present preliminary results concerning precursors to the recent significant (Mw 6.7) earthquake (EQ) that took place in the region south of the Greek Island of Zakynthos (October 25, 2018). We show, in terms of fracture-induced electromagnetic emissions (EME), that the Earth system around the focal area was at critical state up to a few hours before the occurrence of the EQ. Importantly, departure from the critical state, in terms of the symmetry breaking phenomenon, was also identified in the EME recorded after the appearance of the last signatures of criticality. The analysis was performed by means of the method of critical fluctuations (MCF).

**Keywords:** Fracture-induced electromagnetic emissions; Earthquake; Criticality; Symmetry Breaking Phenomenon; Greece.

## 1 Introduction

The possible connection of the electromagnetic (EM) activity that is observed prior to significant earthquakes (EQs) with the corresponding EQ preparation processes, often referred to as seismo-electromagnetics, has been intensively investigated during the last years. Several possible EQ precursors have been suggested in the literature [*Uyeda et al.*, 2009, 2013; *Cicerone et al.*, 2009; *Hayakawa*, 2013a, 2013b, 2015]. The possible relation of the field observed fracture-induced electromagnetic emissions (EME) in the frequency bands of MHz and kHz has been examined in a series of publications [e.g., *Eftaxias et al.*, 2001, 2004, 2008, 2013; *Eftaxias and Potirakis*, 2013; *Contoyiannis et al.,* 2004a, 2005, 2010, 2013, 2015, 2017; *Contoyiannis and Eftaxias, 2008*; *Contoyiannis and Potirakis,* 2018; *Kapiris et al.*, 2004; *Karamanos et al.,* 2006; *Papadimitriou et al.*, 2008; *Potirakis et al.*, 2011, 2012a, 2012b, 2012c, 2013, 2015, 2016; *Minadakis et al.*, 2012a, 2012b; *Donner et al.*, 2015; *Kalimeris et al.*, 2016], while a four-stage model for the preparation of an EQ by means of its observable EM activity has been recently put forward [*Donner et al.*, 2015; *Potirakis et al.*, 2016; *Eftaxias et al.*, 2018, and references therein].

In this letter, we report the recording, with a sampling rate of 1 sample/s, of MHz EME signals with critical characteristics of a second order phase transition in equilibrium before the recent Mw = 6.7 EQ which happened on 25/10/2018 22:54:49 south of the Greek Island of Zakynthos (Zante), epicenter (37.34 °N, 20.51 °E), focal depth 10km (information reported by the National Observatory of Athens, Institute of Geodynamics (NOA-IG) http://bbnet.gein.noa.gr/mt_solution/2018/181025_22_54_49.00_MTsol.html). Moreover, we report that the signature of departure from the critical state in terms of the symmetry breaking phenomenon was identified in the MHz EME recorded right after the last signatures of criticality. This phenomenon signifies the transition from a highly symmetrical state (critical state), to a low symmetry state, during which the process is focused around "preferred" directions. Furthermore, it has to be mentioned that these are preliminary results, while the analysis of the EME recorded at different stations of our network (ELSEM-Net, see Sec. 3) is still in process.

The analysis of the specific EME time series was performed using the method of critical fluctuations (MCF) (see Sec. 2) [*Contoyiannis and Diakonos*, 2000; *Contoyiannis et al.*, 2002, 2013, 2015; *Potirakis et al.*, 2016, 2019]. The analysis reveals that first in the timeline appear critical features in the MHz EME, implying that the possibly related underlying fracture process involved in the preparation of the main shock is at critical state. The presence of the "critical point" during which any two active parts of the system are highly correlated even at arbitrarily long distances, in other words when "everything depends on everything else", is consistent with the view that the EQ preparation process during the period that the MHz EME precursory signals are emitted is a spatially extensive process. It is noted that, according to the aforementioned four-stage model [*Potirakis et al.*, 2016; *Eftaxias et al.*, 2018, and references therein], the pre-seismic critical MHz EM emission is considered to originate during the fracture of the part of the Earth's crust that is characterized by high heterogeneity. During this phase, the fracture is non-directional and spans over a large area that surrounds the family of large high-strength entities (asperities) distributed along the main fault sustaining the system. Note that for an EQ of magnitude ~7 the corresponding fracture process extends to a radius of ~200 km [*Bowman et al.*, 1998]. Thus, during this phase the fracture process is extended up to the land of the neighboring islands and Peloponnese (mainland in the south of Greece). For the EQ of interest, the analysis also reveals that next in the timeline, after the critical MHz EME, appears the symmetry breaking phenomenon in the MHz EME recordings. This finding signifies the departure from critical state through the shrinking of the area in which fractures happen and progressively its restriction around a preferred direction (the fault) [*Potirakis et al.*, 2016; *Eftaxias et al.*, 2018; *Contoyiannis and Potirakis,* 2018].

## 2   Data analysis method

The analysis of the recorded data was performed using the method of critical fluctuations (MCF) [*Contoyiannis and Diakonos*, 2000; *Contoyiannis et al.*, 2002]. Detailed descriptions of all the involved calculations can be found elsewhere [*Contoyiannis et al.*, 2013, 2015; *Potirakis et al.*, 2016, 2019] and therefore are omitted here for the sake of brevity and focus on the findings. However, a general description of the employed method follows.

MCF was proposed for the analysis of critical fluctuations in the observables of systems that undergo a continuous (second-order) phase transition [*Contoyiannis and*

*Diakonos*, 2000; *Contoyiannis et al.*, 2002]. It is based on the finding that the fluctuations of the order parameter, that characterizes successive configurations of critical systems at equilibrium, obey a dynamical law of intermittency of an 1D nonlinear map form. The MCF is applied to stationary time windows (time series excerpts) of statistically adequate length, for which the distribution of the of waiting times $l$ (laminar lengths) of fluctuations in a properly defined laminar region is fitted by a function $f(l) \propto l^{-p_2} e^{-p_3 l}$. The criteria for criticality are $p_2 > 1$ and $p_3 \approx 0$ [*Contoyiannis and Diakonos*, 2000; *Contoyiannis et al.*, 2002], note that a time series excerpt satisfying these criteria is often referred to as "critical window" (CW). In that case the system is characterized by intermittent dynamics, since the distribution follows power-law decay [*Schuster*, 1998]. On the other hand, in the case of a system governed by noncritical dynamics the corresponding distribution follows an exponential decay, rather than a power-law one [*Contoyiannis et al.*, 2004b]. A system in critical dynamics may depart from this high symmetry phase towards a low symmetry (highly localized along a preferred direction) either by the so-called "symmetry-breaking" (SB) phenomenon [*Contoyiannis et al.*, 2004a; *Contoyiannis and Potirakis,* 2018], or by means of a tricritical crossover [*Contoyiannis et al.*, 2015]. The so-called "tricritical point" is the point in the phase diagram of the system at which the two basic kinds of phase transition (second-order phase transition and first-order phase transition) meet. Both these ways of departure from critical state can be detected by using MCF. Symmetry breaking is characterized by marginal presence of power-law distribution, which indicates that the system's state is still close to the critical point. Using MCF, this is identified by the appearance of an extremely narrow range of laminar regions (often just one region) for which criticality conditions are still satisfied, after the appearance of a clear critical window [*Contoyiannis et al.*, 2004a]. Tricritical dynamics are identified in terms of MCF by the appearance of the combination of exponents $p_2 < 1$ and $p_3 \approx 0$ [*Contoyiannis et al.*, 2015]. The MCF has been applied to a variety of dynamical systems, including thermal (e.g., 3D Ising) [*Contoyiannis et al.*, 2002], geophysical [*Contoyiannis et al.*, 2004a, 2005, 2010, 2013, 2015; *Contoyiannis and Eftaxias* 2008; *Potirakis et al.*, 2016; *Contoyiannis and Potirakis,* 2018], biological (electro-cardiac signals) [*Contoyiannis et al.*,2004b; *Contoyiannis et al.*, 2013], economic [*Ozun et al.*, 2014] and electronic systems [*Potirakis et al.*, 2017].

### 3   Analysis results

The hereafter presented signals were recorded in specific remote EME stations of ELSEM-Net (hELlenic Seismo-ElectroMagnetics Network), our ground-based telemetric network spanning across Greece (Fig. 1, Table 1). In the following we present the analysis results for the EME recordings of specific stations of our network. Specifically, those for which the up to now analysis has revealed precursory signs. Our up to now findings show that the evolution of the SB phenomenon, which has been suggested as a precursor signal before a strong seismic events [*Contoyiannis and Potirakis*, 2018], was embedded in the MHz EME recordings of two stations (V, J, see Fig. 1 and Table 1) from ~2 days up to ~12 h before the EQ event.

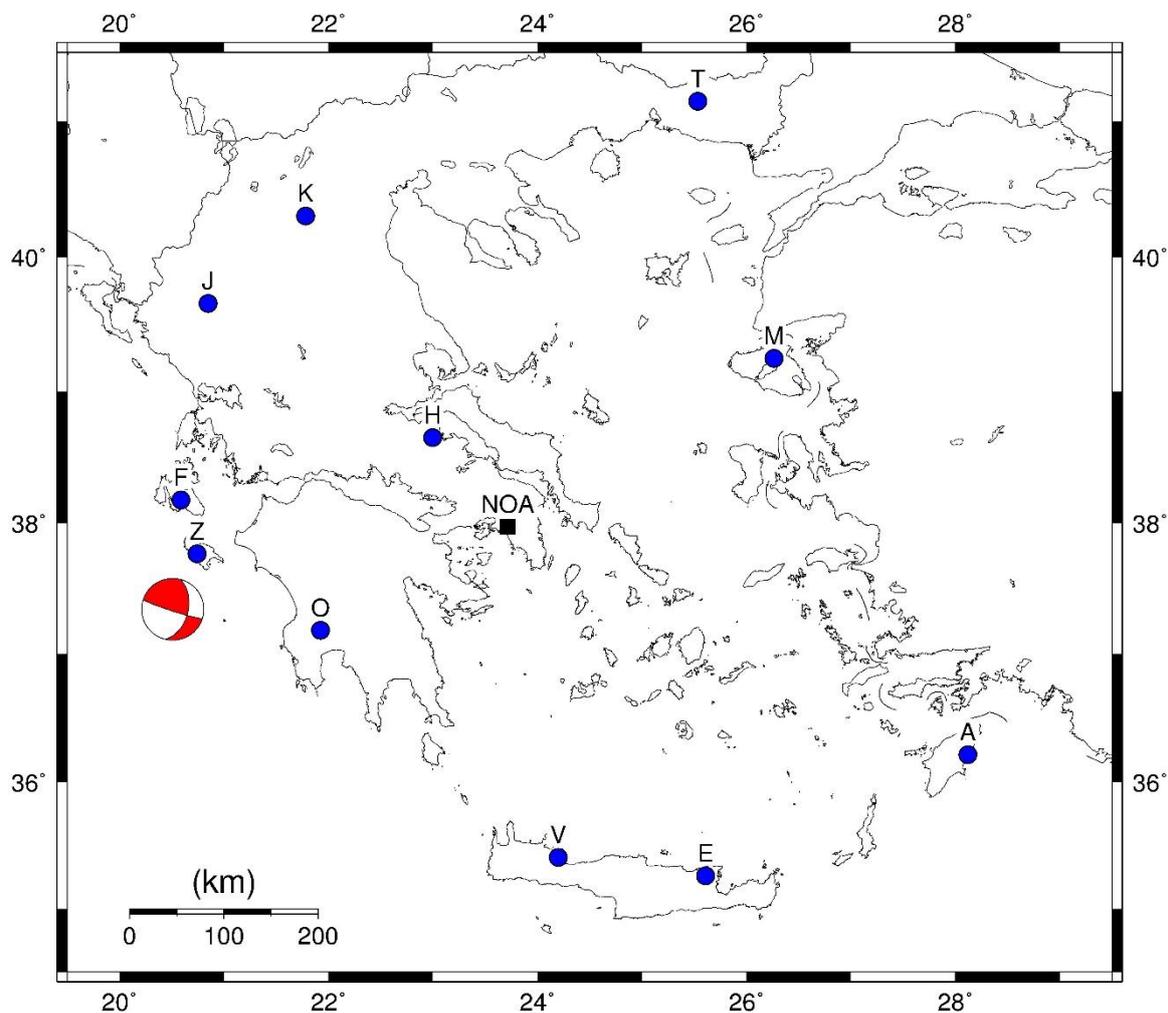

**Fig. 1.** Map showing the 11 stations of the ELSEM-Net, a telemetric network developed for the recording of MHz and kHz EME in Greece. All stations are linked to NOA-IG in Athens. The EQ is noted with the MT solution computed by NOA-IG (see http://bbnet.gein.noa.gr/mt_solution/2018/181025_22_54_49.00_MTsol.html).

**Table1.** Location of the remote EME stations of ELSEM-Net. (* = Currently non-operating)

| No | Station Code | Station Location Name | Latitude (° N) | Longitude (° E) | Altitude (m) |
|---|---|---|---|---|---|
| 1 | J | Ioannina | 39.6561 | 20.8487 | 526 |
| 2 | H | Atalandi | 38.6495 | 22.9988 | 185 |
| 3 | F | Valsamata, Cephalonia Island | 38.1768 | 20.5886 | 402 |
| 4 | O | Ithomi, Mesinia | 37.1787 | 21.9252 | 423 |
| 5 | K | Kozani | 40.3033 | 21.7820 | 791 |
| 6 | E | Neapoli, Crete Island | 35.2613 | 25.6103 | 288 |
| 7 | V | Vamos, Crete Island | 35.4070 | 24.1997 | 225 |
| 8 | A | Archangelos, Rhodes Island | 36.2135 | 28.1212 | 148 |
| 9 | T | Komotini | 41.1450 | 25.5355 | 116 |
| 10 | M | Agia Paraskevi, Lesvos Island | 39.2456 | 26.2649 | 130 |
| 11 | Z (*) | Fterini-Aghios Leon, Zakynthos Island | 37.7658 | 20.7430 | 461 |

The SB phenomenon evolution is embedded in the MHz time series recorded by Vamos (V) and Ioannina (J) stations from ~2 days to ~ 12 h before the EQ event, as shown in Figs. 2 and 3, respectively. The colors correspond to the following sequence of states of the underlying system which describe the evolution of the SB phenomenon: critical state (green), symmetry breaking (red), departure from critical state / noncritical dynamics (blue, magenta). Interestingly, we observe that the time interval for which the specific sequence was revealed for the Ioannina station is entirely within the corresponding interval of Vamos station, indicating practically a coincidence of the observed evolution in these two stations. The analysis of each segment of these two time series is presented in Figs. 4-7 for the Vamos signal and in Figs. 8-10 for the Ioannina signal, respectively.

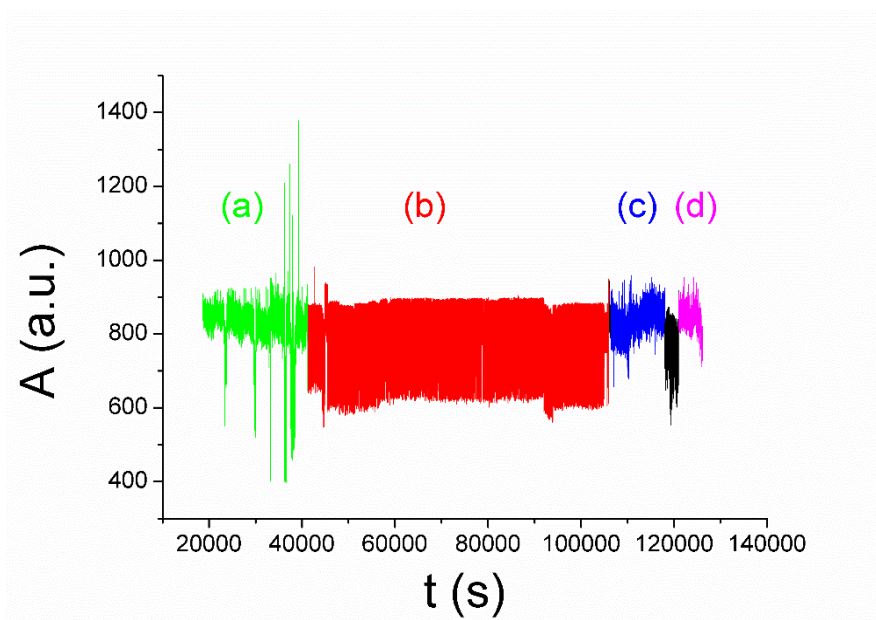

**Fig. 2.** The evolution of the SB phenomenon as embedded in the recordings of the Vamos station: critical state (a), symmetry breaking (b), departure from critical state (noncritical dynamics) (c), further departure from critical state (noncritical dynamics) (d). Time in sec from 00:00:00 UT of 24/10/2018, i.e., spans from 24/10/2018 05:11:40 UT to 25/10/2018 11:03:20 UT.

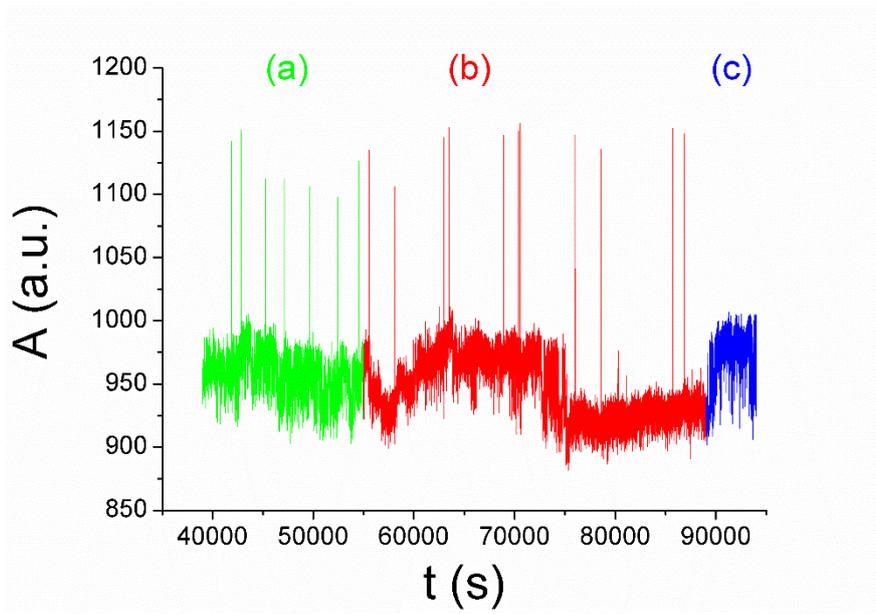

**Fig. 3.** The evolution of the SB phenomenon as embedded in the recordings of the Ioannina station: critical state (a), symmetry breaking (b), departure from critical state (noncritical dynamics) (c). Time in sec from 00:00:00 UT of 24/10/2018, i.e., spans from 24/10/2018 10:50:00 UT to 25/10/2018 02:06:40 UT.

First we present in detail the analysis of the critical window of the Vamos station signal (Fig. 2a, Fig. 4a). The main steps of the MCF analysis [e.g., *Contoyiannis et al.*, 2013; *Potirakis et al.*, 2019] on the critical window of Fig. 4a are shown in Figs. 4b-d. First, a distribution of the amplitude values of the analyzed signal was obtained (Fig. 4b) from which, using the method of turning points [*Pingel et al.*, 1999], a fixed-point, that is the start of laminar regions, $\phi_o$ was determined. Fig. 4c portrays an example of the obtained laminar distributions for a specific end point $\phi_l$, that is the distribution of waiting times, referred to as laminar lengths $l$, between the fixed-point $\phi_o$ and the end point $\phi_l$, as well as the fitted function $f(l) \propto l^{-p_2} e^{-p_3 l}$. Finally, Fig. 4d shows the obtained plot of the $p_2$, $p_3$ exponents vs. $\phi_l$. From Fig. 4d it is apparent that the criticality conditions, $p_2 > 1$ and $p_3 \approx 0$, are satisfied for a wide range of end points $\phi_l$, revealing the power-law decay feature of the time series which indicates that the system is characterized by intermittent dynamics; in other words, the MHz time series excerpt of Fig. 4a is indeed a CW.

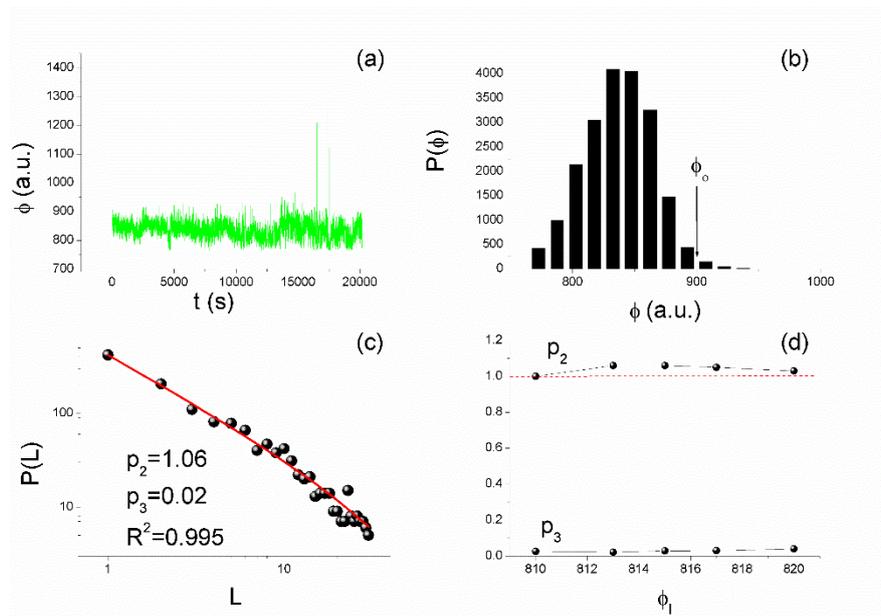

**Fig. 4**. MCF analysis of the time series excerpt of Fig. 2a. (a) The 20000 samples long critical window of the MHz EME that was recorded before the 2018 Zakynthos EQ at the Vamos station (see also Fig. 1, Table 1). (b) Amplitude distribution of the signal of (a). (c) A representative example of laminar distribution and the involved fitting. The solid line corresponds to the fitted function (cf. to text in Sec. 2). (d) The obtained exponents $p_2$, $p_3$ vs. different values of the end of laminar region $\phi_l$. The horizontal dashed line indicates the critical limit ($p_2 = 1$). Critical behavior ($p_2 > 1$ and $p_3 \approx 0$) is obvious.

The second time series excerpt of the Vamos signal (Fig. 2b) has a values' histogram which presents two lobes as depicted in Fig. 5, indicating the SB phenomenon according to theory of critical phenomena [*Huang*, 1987] and the results of a recent of a work of ours [*Contoyiannis and Potirakis,* 2018] under the condition that the next window is not a critical window. Indeed, the analysis of two following excerpts of Vamos signal (Fig. 2c and 2d), which is presented in Figs. 6 and 7, respectively, indicates that the underlying system has departed from critical state, with the exponential term of the function $f(l) \propto l^{-p_2} e^{-p_3 l}$ (see Sec. 2) progressively prevailing.

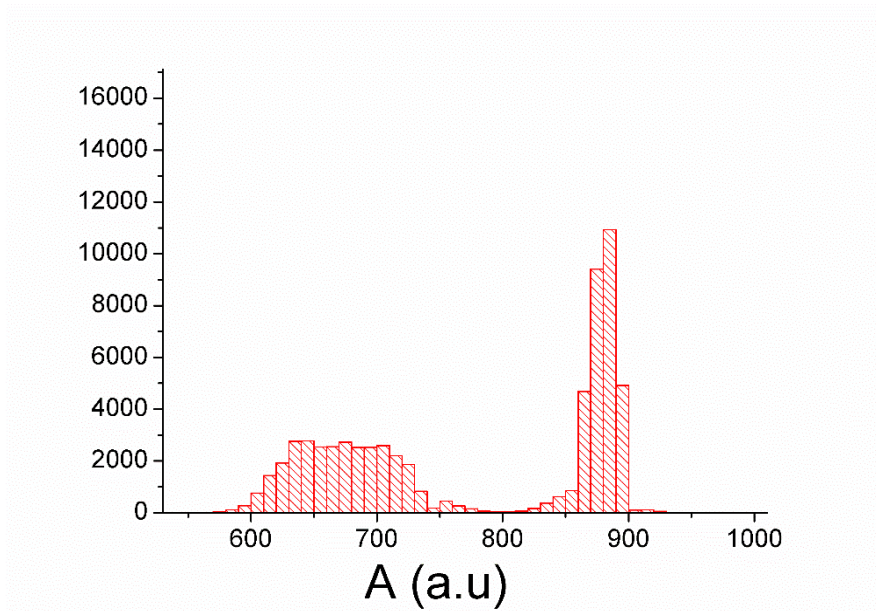

**Fig. 5**. Values' histogram of the time series excerpt of Fig. 2b.

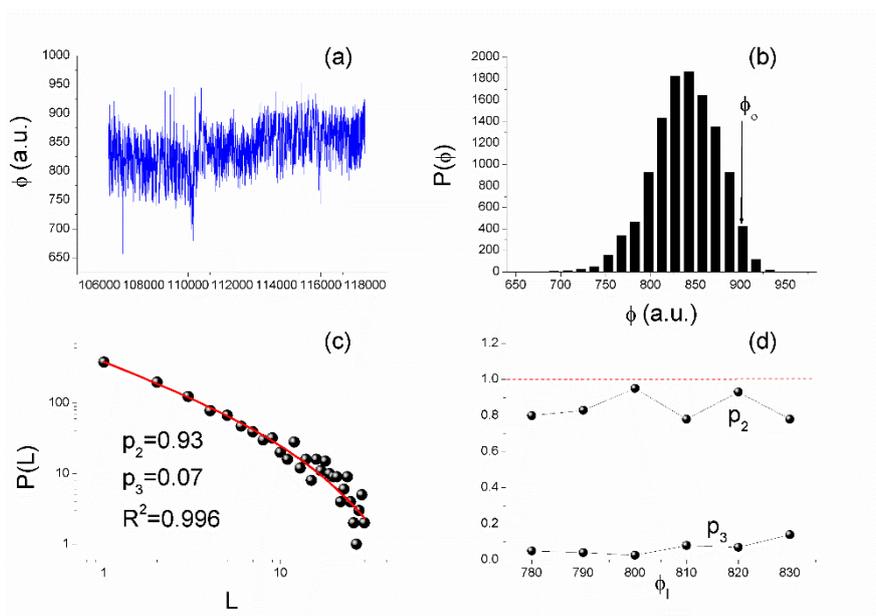

**Fig. 6**. MCF analysis of the time series excerpt of Fig. 2c. Figure structure follows that of Fig. 4.

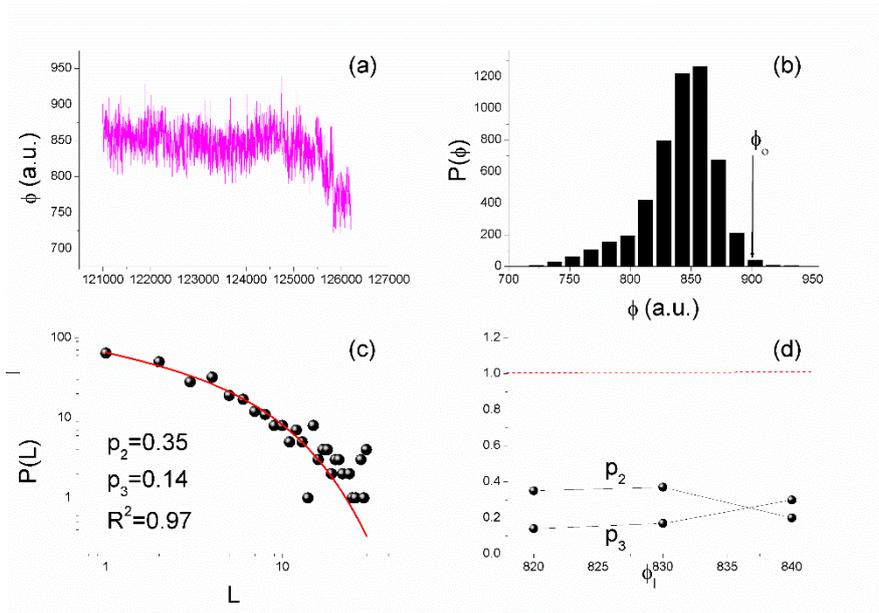

**Fig. 7**. MCF analysis of the time series excerpt of Fig. 2d. Figure structure follows that of Fig. 4.

A similar situation is observed in the consecutive parts of the Ioannina station signal. Specifically, the first time series excerpt (Fig. 3a) presents critical behavior as shown in Fig. 8, the second one (Fig. 3b) indicates SB as depicted in Fig. 9, which is confirmed by the fact that the last one (Fig. 3c) possesses no indication of critical state as shown in Fig. 10.

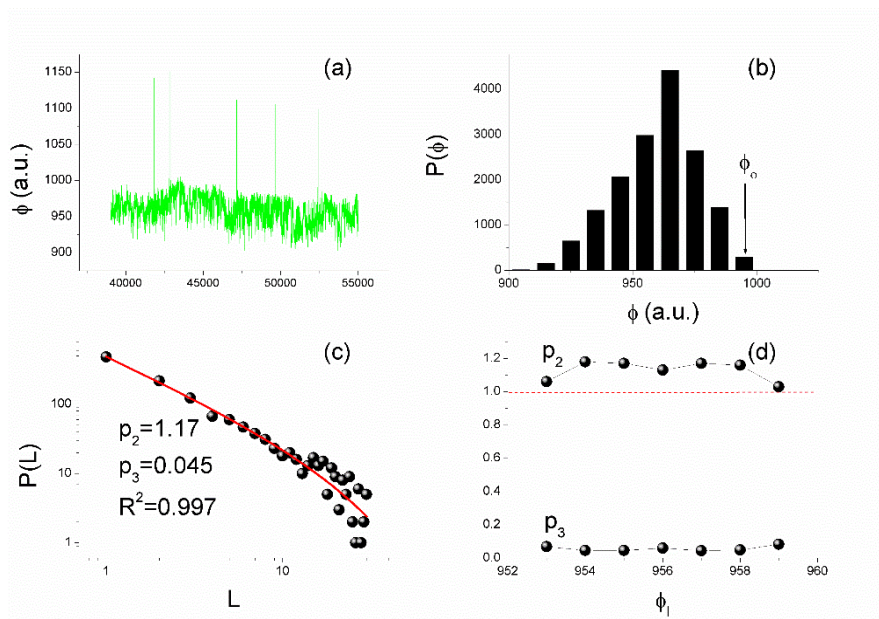

**Fig. 8**. MCF analysis of the time series excerpt of Fig. 3a. Figure structure follows that of Fig. 4.

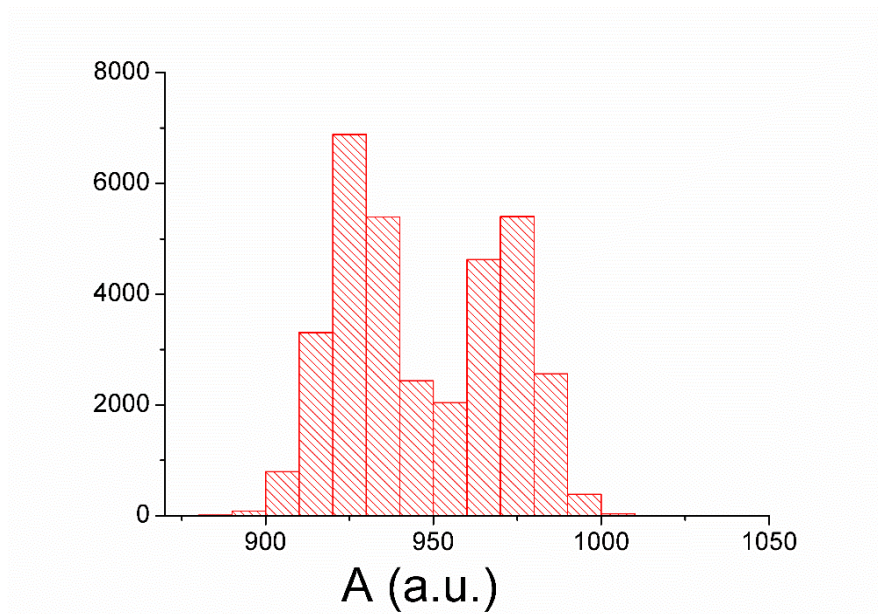

**Fig. 9**. Values' histogram of the time series excerpt of Fig. 3b.

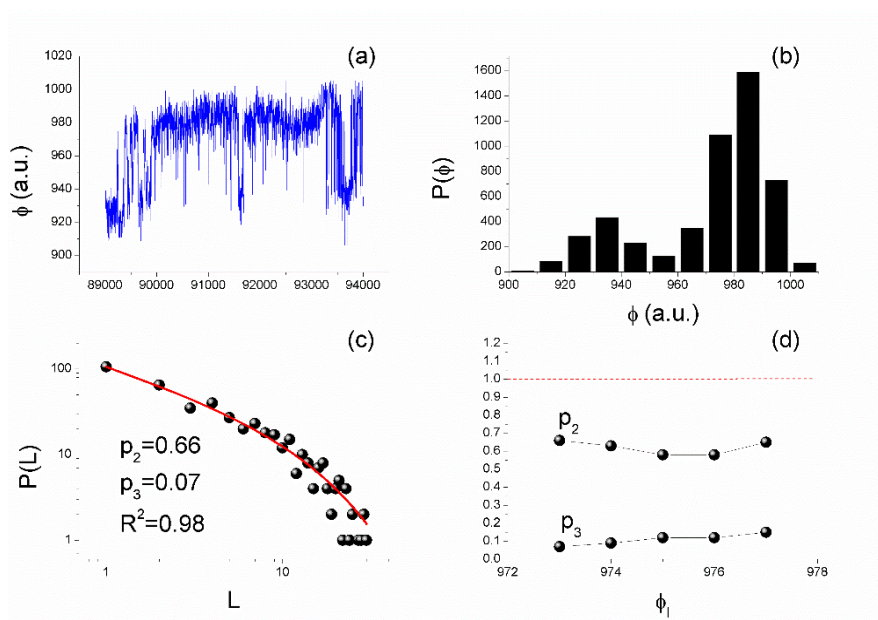

**Fig. 10**. MCF analysis of the time series excerpt of Fig. 3a. Figure structure follows that of Fig. 4.

## 4  Conclusions

Based on the method of critical fluctuations, we have shown that the preparation process of the recent seismic event that occurred on 25 October 2018 in Zakynthos (Greece) has been imprinted in the MHz EME recorded at two EME stations of ground-based telemetric network (ELSEM-Net). Specifically, both Vamos and Ioannina stations, almost simultaneously, recorded a full sequence of the evolution of the symmetry breaking phenomenon from ~2 days to 12 h prior to the EQ. Such a sequence signifies the departure from critical state through the shrinking of the area in which fractures happen and progressively its restriction around a preferred direction (the fault).

It has to be mentioned that indications of critical fluctuations have been found in the MHz EME recordings of different stations of our network even up to 9 days before the EQ, which is agreement with the recently published findings of Sarlis and Skordas [*Sarlis and Skordas*, 2018] that foreshock seismicity reached criticality conditions according to the natural time method on 18 October 2018, following the critical conditions revealed in SES signals on 2 October 2018.

The analysis of the EME recordings of our network is still in process and when this is finished, the results will be presented in a full research article. Finally, it has to be mentioned that, unfortunately, the Zakynthos Island ELSEM-Net station, which is known for its remarkable sensitivity to kHz EME, has been out of service since long period before the occurrence of the EQ.